# Superconducting Material Diagnostics using a Scanning Near-Field Microwave Microscope

Steven M. Anlage, D. E. Steinhauer, C. P. Vlahacos, B. J. Feenstra, A. S. Thanawalla, Wensheng Hu, Sudeep K. Dutta, and F. C. Wellstood

Center for Superconductivity Research, Physics Department, University of Maryland, College Park, MD 20742-4111

**Abstract -- We have developed scanning near-field microwave microscopes which can image electrodynamic properties of superconducting materials on length scales down to about 2 μm. The microscopes are capable of quantitative imaging of sheet resistance of thin films, and surface topography. We demonstrate the utility of the microscopes through images of the sheet resistance of a $YBa_2Cu_3O_{7-\delta}$ thin film wafer, images of bulk Nb surfaces, and spatially resolved measurements of $T_c$ of a $YBa_2Cu_3O_{7-\delta}$ thin film. We also discuss some of the limitations of the microscope and conclude with a summary of its present capabilities.**

## I. Introduction

The ability to fabricate high quality thin films of superconducting materials is the first step in their application to demanding commercial microwave technologies. To be cost-effective, the films should be homogeneous over a wafer, so that devices built on different parts of the wafer will behave as designed. Also, non-linearity, intermodulation and power dependence are problems of crucial importance in applying high-$T_c$ superconductors to microwave communications circuits. Innovative approaches are required to locally characterize superconducting properties and study the causes of power dependence and nonlinearity in superconducting microwave devices. It is particularly important to characterize these materials at the frequencies and temperatures of interest, namely 0.8 - 10 GHz and 77 K. It is also important to examine the microwave properties of materials on a variety of length scales: from the wafer level for homogeneity characterization, down to the sub-micron level for defect characterization. In this paper we discuss the development of near-field microscopy techniques suitable for superconducting material diagnostics.

## II. Principle Of Operation

Scanning microwave microscopes have been used to image materials at microwave frequencies, and to measure the local resistivity variations of semiconducting materials [1]-[5].

Fig. 1 shows a schematic illustration of our apparatus [6]. The microwave signal from the source enters a coaxial transmission line resonator defined on one end by a decoupling capacitor, and on the other by an open-ended coaxial probe. Microwave power is reflected from the open end many times (Q ~ $10^2$ - $10^3$), eventually escaping from the resonator, and is detected by a diode.

For close separation between probe and sample, the reflected signal is strongly influenced by the presence of the sample. The coupling of the system to the sample is predominantly capacitive. If the sample is metallic, it forms one plate of a capacitor, while the other plate is formed by the center conductor of the coaxial probe. As the probe/sample separation decreases, the capacitance $C_x$ increases (see Fig. 1 inset), resulting in a drop of the resonant frequency of the coaxial resonator. In one extreme, when the probe is very far from the sample, the transmission line is open-ended and has only a terminating impedance due to fringe capacitance. In this case, the system is a half-wave resonator with a resonant frequency $f_0$. In the other extreme, when the metallic sample is in contact with the probe, there is a short circuit termination, otherwise known as a Corbino contact [7]. In this case, the system is a quarter-wave resonator, and the resonant frequency is reduced by $\Delta f_{max}$ = (1/2) (c/2L)/$\varepsilon_r^{1/2}$, where c is the speed of light, L is the length of the coaxial resonator, and $\varepsilon_r$ is the relative dielectric constant of the coaxial cable. For intermediate cases of finite probe/sample separation, the frequency shift has values between 0 and $\Delta f_{max}$.

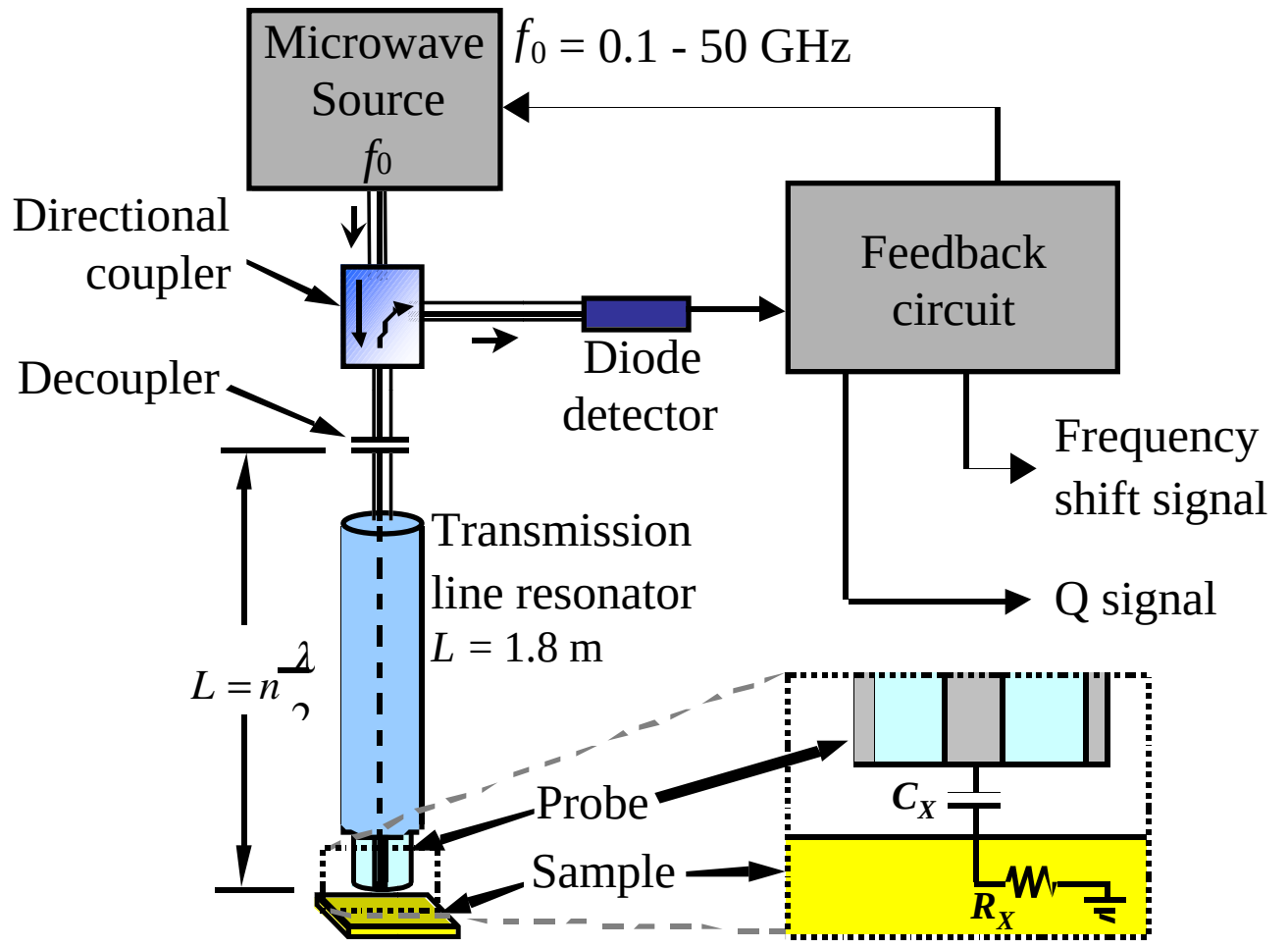

Fig. 1. Schematic diagram of the materials diagnostic scanning near-field microwave microscope. The inset shows the interaction between the open-ended coaxial probe and the sample.

In general, a sample will also cause additional dissipation due to induced currents which flow in the sample, as well as enhanced radiation losses. The increased dissipation can be modeled as a series resistance, $R_x$ between the probe/sample capacitor and ground (see Fig. 1 inset). The resistance has the effect of decreasing the quality factor Q of the coaxial resonator.

To acquire an image, the sample is scanned beneath the coaxial probe. As it moves, the capacitive gap and the local sheet resistance will change, changing the resonant frequency and quality factor of the resonator. To record these changes, we use a feedback loop (Fig. 1) which keeps the microwave source locked to a particular resonant frequency of the coaxial

Manuscript received September 15, 1998.
This work was supported by the National Science Foundation through grants NSF ECS-9632811 and NSF-MRSEC DMR-9632521, and the Maryland Center for Superconductivity Research.

resonator [8]. This feedback loop operates by frequency modulating the source at a frequency $f_{FM}$ = 3 kHz, and zeroing out the first harmonic response of the system. With the loop in operation, we simply record the feedback signal as a function of position. The second harmonic ($2f_{FM}$) signal ($V_{2f}$) on resonance is related to the Q of the resonator, and can be recorded simultaneously. We have also recorded the amplitude of the reflected signal on resonance, but find that it gives essentially the same information as that contained in the Q image.

We currently have two room temperature and one cryogenic microwave microscope. One of the room temperature microscopes is shown in Fig. 2. The sample is supported by a manual z-axis translation stage and a two-axis leveling assembly. This, in turn, is mounted on a motorized x-y translation stage with a 0.1 µm step resolution. The probe is held above the sample on a motorized z-axis translation stage, fixed to a rigid frame. A digital camera and binocular microscope are used to monitor the probe/sample separation during the scan. The microscope shown in Fig. 2 is located on a vibration isolation table, which allows imaging with sub-micron resolution.

The cryogenic microscope is based on the design of scanning SQUID microscopes developed in our laboratory. It consists of a cryogenic xy-slider system which translates the sample with approximately 1 µm resolution, and a variable height (z) coaxial probe holder. The xy-motion is achieved by computer-controlled stepper motors at room temperature which turn vacuum feedthroughs connected to the cryogenic slider through thin-walled stainless steel tubes. The microscope can operate at temperatures between 4.2 K and room temperature.

## III. MICROSCOPE CHARACTERISTICS

### *A. Spatial Resolution*

The spatial resolution of the microscope is dictated by two length scales which are under our control. The first is the length scale of the enhanced electromagnetic fields associated with the probe [9]. In our case, the probe size is given by the diameter of the center conductor of the coaxial probe, or by the radius of curvature of a "lightning rod" feature of the probe. By substituting a sharp STM tip for the center conductor, and thus localizing more strongly the electromagnetic fields, we can reduce this length scale significantly.

The second important length scale is the probe/sample separation. In our systems, the separation is much less than the free space wavelength of the radiation. However, to obtain the ultimate spatial resolution, the probe must be closer than the probe size mentioned above. In principle, it is possible to establish STM tunneling into conducting samples to maintain a fixed height [10], or to use an AFM in contact mode [11].

The best resolution we have obtained is for operation in contact with an STM tip substituting for the center conductor of the coaxial probe. The original center conductor is removed, replaced by a hypodermic tube, and an STM tip is placed inside (see Fig. 3(a)) [10]. Using this configuration we imaged a series of 0.5 µm wide Al thin film lines deposited on mylar (see photograph in Fig. 3(b)) at 7.5 GHz. The lines are spaced 2 µm apart. Fig. 3(c) shows the frequency shift scan along the dashed line shown in Fig. 3(b). The large drops in frequency coincide with the Al lines, and have a 2 µm periodicity. The variation in the signal strength at each metallic line is due to intermittent contact as the probe is dragged across the sample. This demonstrates that in this configuration, the microwave microscope has a spatial resolution of better than 2 µm.

One of the advantages of our microscopes is the wide dynamic range of spatial resolution. We can image an entire wafer using a coaxial probe with an inner conductor diameter of 480 µm (as shown below), or on very fine scales using an STM-tipped coaxial probe. Other probe sizes we have employed include 200 µm, 100 µm, and 12 µm center conductor diameters [9].

### *B. Frequency Bandwidth*

The spatial resolution of the microscope is independent of the measurement frequency and the system operates over a wide bandwidth. This means that images of material properties can be obtained at precisely the frequency at which the materials will be used. For example, consider a microscope with a coaxial resonator length L = 2 m. In this case, the fundamental mode will be at approximately 50 MHz, and there will be overtones available for imaging at all integer multiples of 50 MHz. The upper frequency limit of the microscope is set by the bandwidth of the electronics. In most practical situations, the microwave source sets the upper frequency limit, about 50 GHz in our case. However, the microwave directional coupler, detector, and coaxial cable connectors can also limit the bandwidth of the microscope. Nevertheless, it is possible to construct a microscope which has near-continuous imaging capability over three decades of frequency between 50 MHz and 50 GHz.

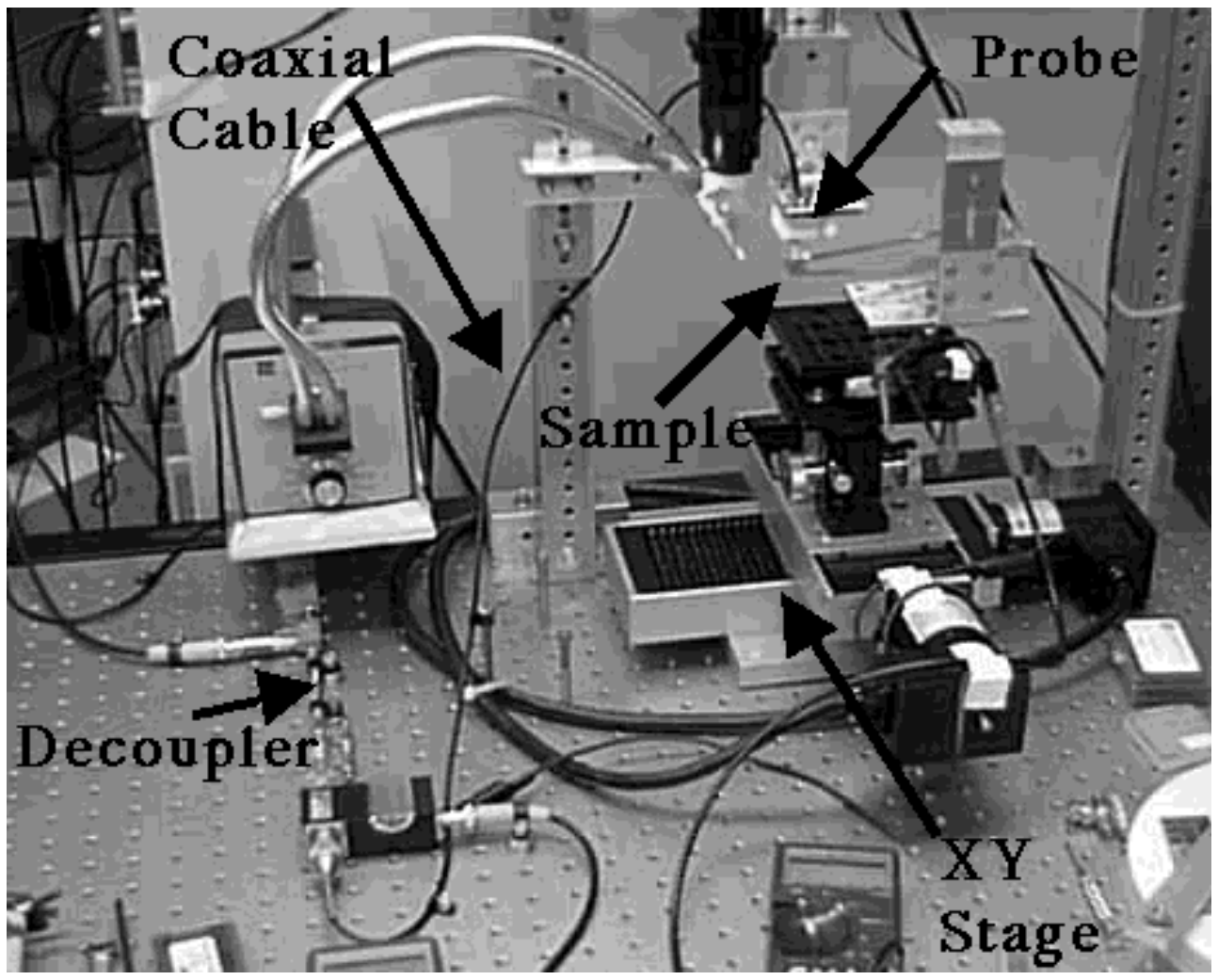

Fig. 2. Room temperature scanning near-field microwave microscope. The sample is located on the xyz-translation stage at the right, and the probe is on a z-axis translation stage suspended from the frame. The coaxial resonator is defined by a decoupler on the optical table.

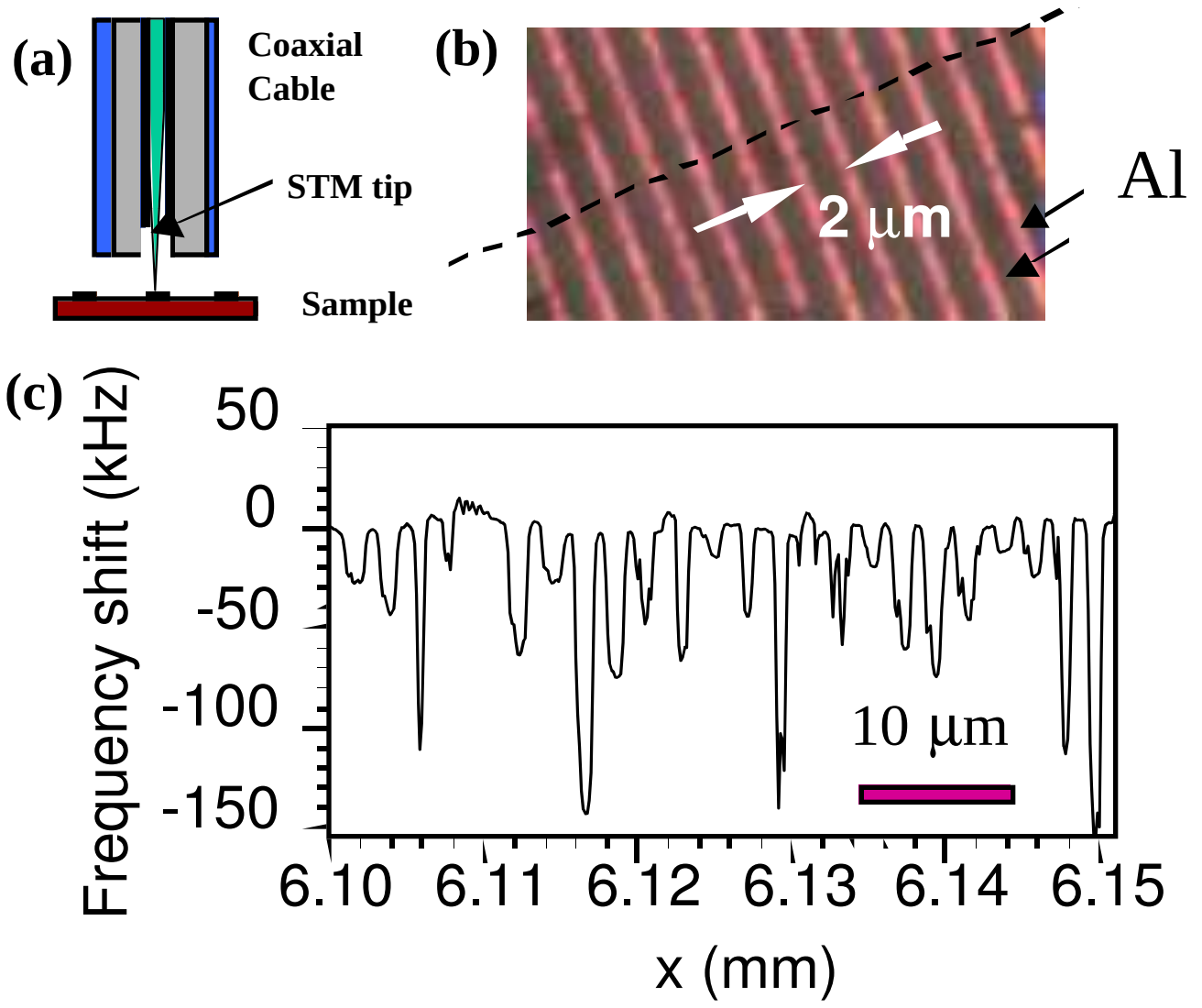

Fig. 3. a) Schematic of STM-tipped coaxial probe. b) Optical micrograph of infrared polarizer grid consisting of Al thin film lines on mylar. c) Frequency shift vs. position for the scan along the dashed line in b).

## IV. IMAGES

### *A. Sheet Resistance of $YBa_2Cu_3O_{7-\delta}$ Thin Film*

Non-destructive imaging of microwave sheet resistance has been demonstrated using a variety of resonant probe systems. For non-destructive sheet resistance imaging of thin films, it is desirable to have quantitative methods that combine high resolution, high speed, simple construction from commercially-available components, and straightforward image interpretation.

Dissipation in a sample is best imaged through measurements of the Q [12]. To determine the relationship between the microscope Q and sample sheet resistance ($R_x$), we used a variable-thickness aluminum thin film on a glass substrate [8]. The cross-section of the thin film is wedge-shaped, producing a spatially varying sheet resistance. Using a probe with a 500 µm diameter center conductor, and selecting a resonance of the microscope with a frequency of 7.5 GHz, we acquired frequency-shift and Q data. We then cut the sample into narrow strips to take two-point resistance measurements and determine the local sheet resistance. We found that the microscope Q reaches a maximum as $R_x$ -> 0; as $R_x$ increases, Q drops due to loss from currents induced in the sample, reaching a minimum around $R_x$ = 660 Ω/sq. for a height of 50 µm. Similarly, as $R_x$ -> ∞, Q increases due to diminishing currents in the sample [12].

To explore the capabilities of our system, we scanned a thin film of $YBa_2Cu_3O_{7-\delta}$ (YBCO) on a 5 cm-diameter sapphire substrate at room temperature. The film was deposited using pulsed laser deposition with the sample temperature controlled by radiant heating. The sample was rotated about its center during deposition, with the ~3 cm diameter plume held at a position halfway between the center and the edge. The thickness of the YBCO thin film varied from about 100 nm at the edge to 200 nm near the center.

Frequency shift and Q data were acquired simultaneously, using a probe with a 500 µm-diameter center conductor at a height of 50 µm above the sample, and a frequency of 7.5 GHz. The scan took approximately 10 minutes to complete, with raster lines 0.5 mm apart. We then transformed the Q data to the sheet resistance image in Fig. 4 using the Q vs $R_x$ calibration data at a height of 50 µm [12]. Fig. 4 confirms that the film does indeed have a lower resistance near the center, as was intended when the film was deposited. We note that the sheet resistance does not have a simple radial dependence, due to either non-stoichiometry or defects in the film.

After scanning the YBCO film, we patterned it and made four-point dc resistance measurements all over the wafer (see Fig. 5). The dc sheet resistance had a spatial dependence identical to the microwave data in Fig. 4. However, the absolute values were approximately twice as large as the microwave results, probably due to degradation of the film during patterning.

### *B. Topography*

The microwave microscope is quite sensitive to variations in probe/sample separation. For a uniformly conducting sample, the height variations simply change the capacitive coupling between the probe and the sample, which changes the resonant frequency of the microscope. These changes in resonant frequency vs. height can be calibrated, and this calibration can be applied to quantitatively determine the height of a uniformly conducting sample as a function of lateral position. We have found that this quantitative topographic imaging capability can achieve height sensitivity as good as 55 nm at a nominal probe/sample separation of 30 μm [13],[14].

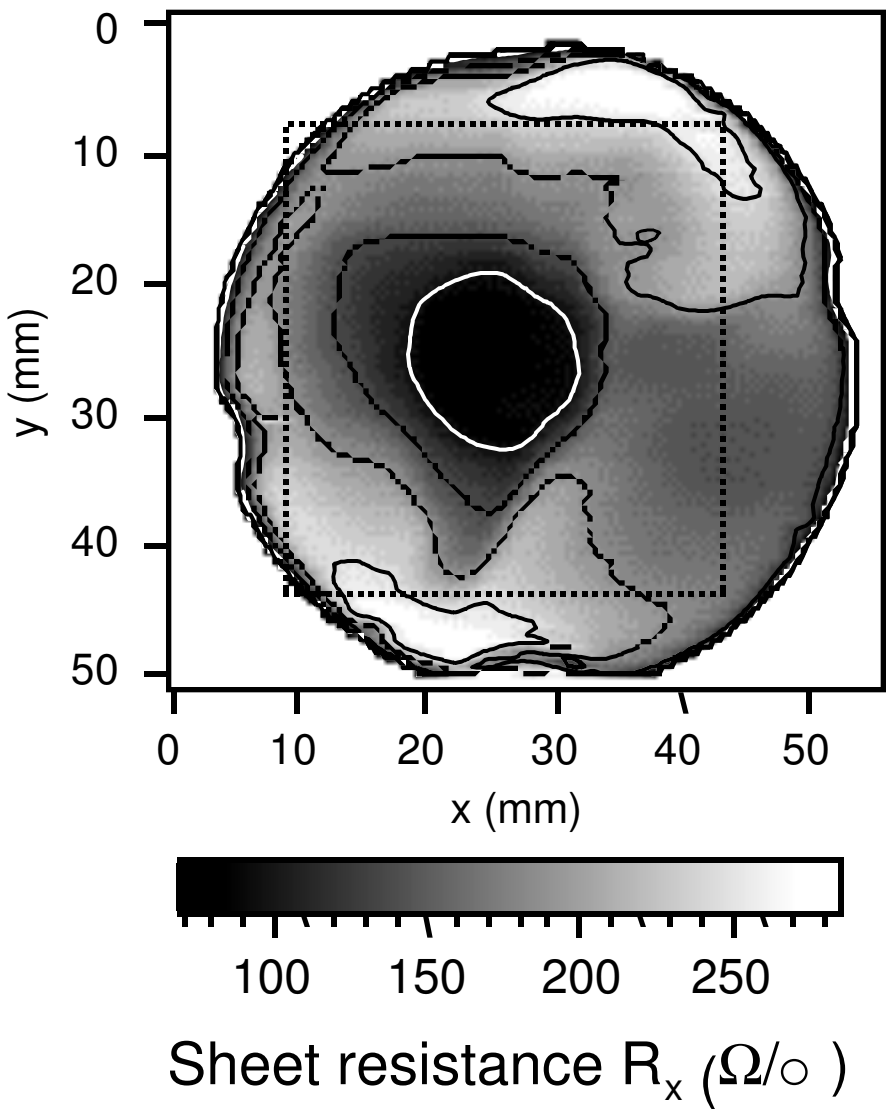

Fig. 4. Room temperature image of the sheet resistance of a YBCO thin film on a 5 cm diameter sapphire wafer. A probe with a 500 µm diameter center conductor was used at a height of 50 µm and a frequency of 7.5 GHz. The contour lines (starting from the center) are at 100, 150, 200, and 250 Ω/square. The dc sheet resistance in the dotted square is shown in Fig. 5.

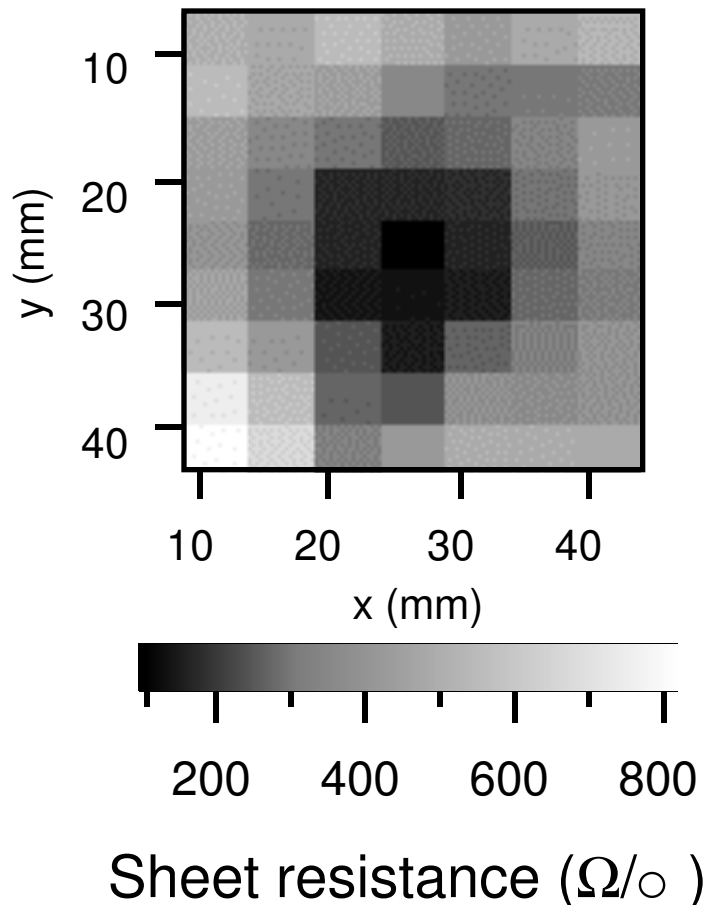

Fig. 5. Map of the dc sheet resistance of the patterned YBCO thin film shown in Fig. 4.

### *C. Surface Imaging of Bulk Nb*

We have used the near-field scanning microwave microscope to image the high frequency properties of a bulk Nb surface. A Nb heat treatment was performed by Peter Kneisel of the Thomas Jefferson National Accelerator Laboratory, and the sample was subsequently etched several times with an aggressive $HF/HNO_3$ acid treatment. Optical micrographs of the surface show irregular grains with sizes ranging from 400 μm to about 1.5 mm in diameter, with most on the order of 1 mm in diameter. Upon closer examination, within the grains we observe surface structures which are approximately 10 μm across. On some grains these structures are pyramidal in shape, while in others they are dome-like, resembling bubbles. The grain boundaries were also deeply etched and have a total width of approximately 10 μm on the surface.

We imaged the bulk Nb surfaces at room temperature with a 480 μm inner conductor diameter probe (see Fig. 6). Both the Q and frequency shift images show a cleared hole in the Nb piece, as well as a scratch on the surface approximately 3 mm long and several hundred microns deep. Both features are easily imaged and produce excellent contrast.

We next imaged a "featureless" 2 mm x 2 mm area of the Nb surface in the upper right hand corner of Fig. 6. The near-field microwave microscope images shown in Fig. 7 were obtained with a 200 μm inner conductor diameter probe operating at a height of approximately 5 μm and a frequency of 7.47 GHz. The 2f (related to the resonator Q) and frequency shift images are quite similar, showing common features which are on the scale of the probe inner conductor diameter, or greater. The 2f image shows regions of low Q on length scales less than the typical grain size, which may be associated with localized areas having a lateral extent less than 200 μm. The 2f image also shows larger areas, on the order of a typical grain size, with lower Q. The frequency shift image reproduces many of the features seen in the 2f image.

It should be noted that the dark regions in Fig. 7 show both a decrease in Q and a downward frequency shift. There are two scenarios for the surface features which produce this kind of image. The first scenario is that the surface has a uniform surface resistance and variable surface topography [13]. The dark areas in Fig. 7 would then be interpreted as raised features on the surface. A smaller probe/sample separation produces a downward shift in resonant frequency and Q, independent of surface resistance [8]. In this interpretation, Fig. 7 simply reflects the topography of a more-or-less uniformly conducting surface.

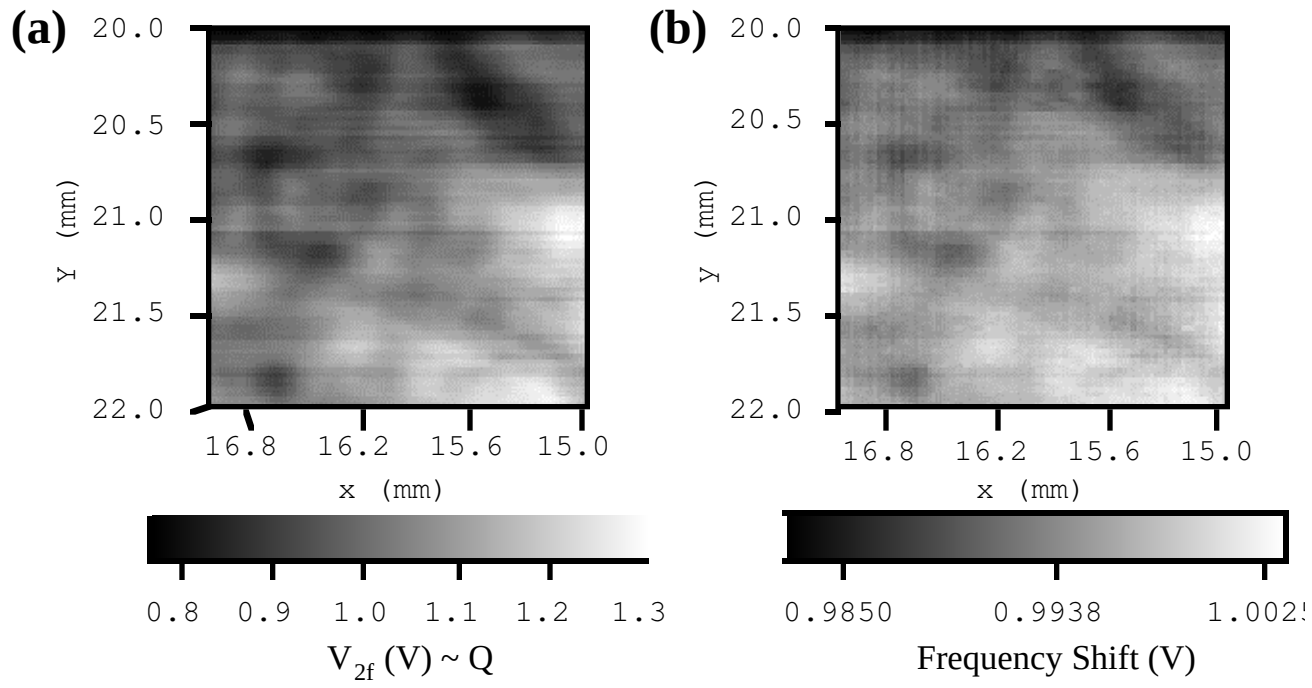

Fig. 7. Microwave microscope images of the a) 2f signal, and b) frequency shift signal for the surface of bulk Nb. These scans cover an area of 2 mm x 2 mm.

The second scenario is that the surface has a flat topography with a variable surface resistance. The dark features in Fig. 7 would then correspond to regions of lower surface resistance, since the frequency shift monotonically becomes more negative as $R_x$ decreases at fixed height. To be consistent with the known Q vs. $R_x$ curve for this probe [8],[12], the surface sheet resistance must exceed approximately 400 Ω/sq. Only in this range of sheet resistance, does the Q decrease with decreasing sheet resistance [8]. In this interpretation the entire Nb surface must have a sheet resistance considerably greater than 400 Ω/sq, with pockets of somewhat lower sheet resistance. However, since the sheet resistance of bulk Nb is expected to be on the order of $10^{-3}$ Ω/sq. at this frequency, the topographic interpretation of the images is almost certainly the correct one.

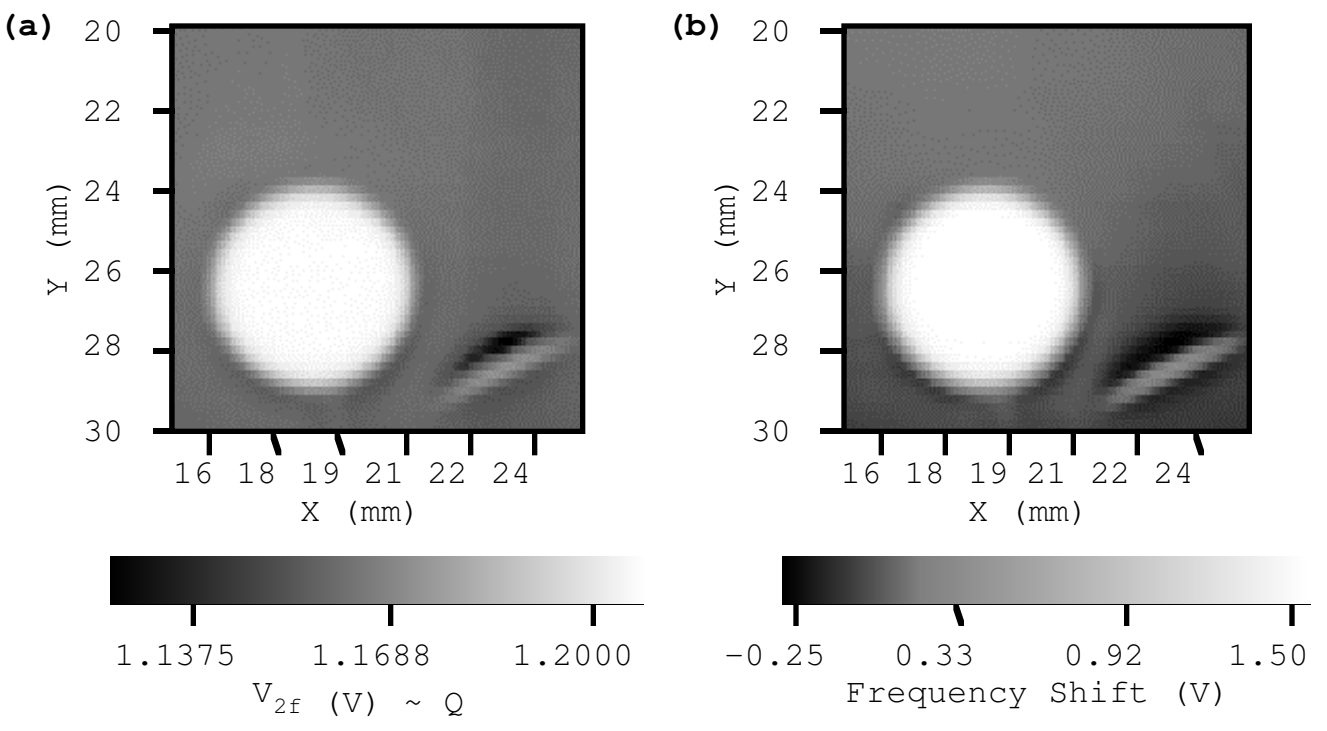

Fig. 6. a) $V_{2f}$ and b) frequency shift images of a cleared hole and 3 mm long scratch in the surface of a bulk Nb sample. The images were obtained with a 480 μm inner conductor diameter probe at a frequency of 7.46 GHz and a height of approximately 50 μm, and cover a 10 mm x 10 mm area. The dark areas correspond to lower Q and more negative frequency shift in the left and right images, respectively.

### *D. Spatially Resolved $T_c$ Imaging*

Although quantitative measurements of the surface impedance are inhibited in the superconducting state due to sensitivity limitations of our current microscope, we can learn much about the quality of a high-$T_c$ thin film, by measuring the local homogeneity of the transition temperature $T_c$. The onset of superconductivity strongly modifies the surface impedance $Z_s$, due to the presence of the superfluid, which dominates the screening behavior below $T_c$. In addition, for high-$T_c$ superconductors, the normal state contribution to the dielectric function changes considerably, due to the rapid change in the quasiparticle scattering rate just below $T_c$. This allows us to study the superconducting transition qualitatively, in order to determine the quality and homogeneity of a particular thin film.

In Fig. 8 we have plotted the temperature dependence of Δf and Q measured at several positions on a 6 mm x 6 mm $YBa_2Cu_3O_{7-\delta}$ thin film. The film was deposited on a $LaAlO_3$ substrate and has a thickness of 200 nm. The measurements were performed in the cryogenic near-field microwave microscope described above.

Both sets of curves were measured at 8.3 GHz while ramping the temperature, keeping a 200 μm diameter probe at a fixed position and a height of 60μm above the sample. The positions on the sample have been indicated by their corresponding symbols in the inset to Fig. 8a. The curves in both panels have been displaced vertically for the sake of clarity.

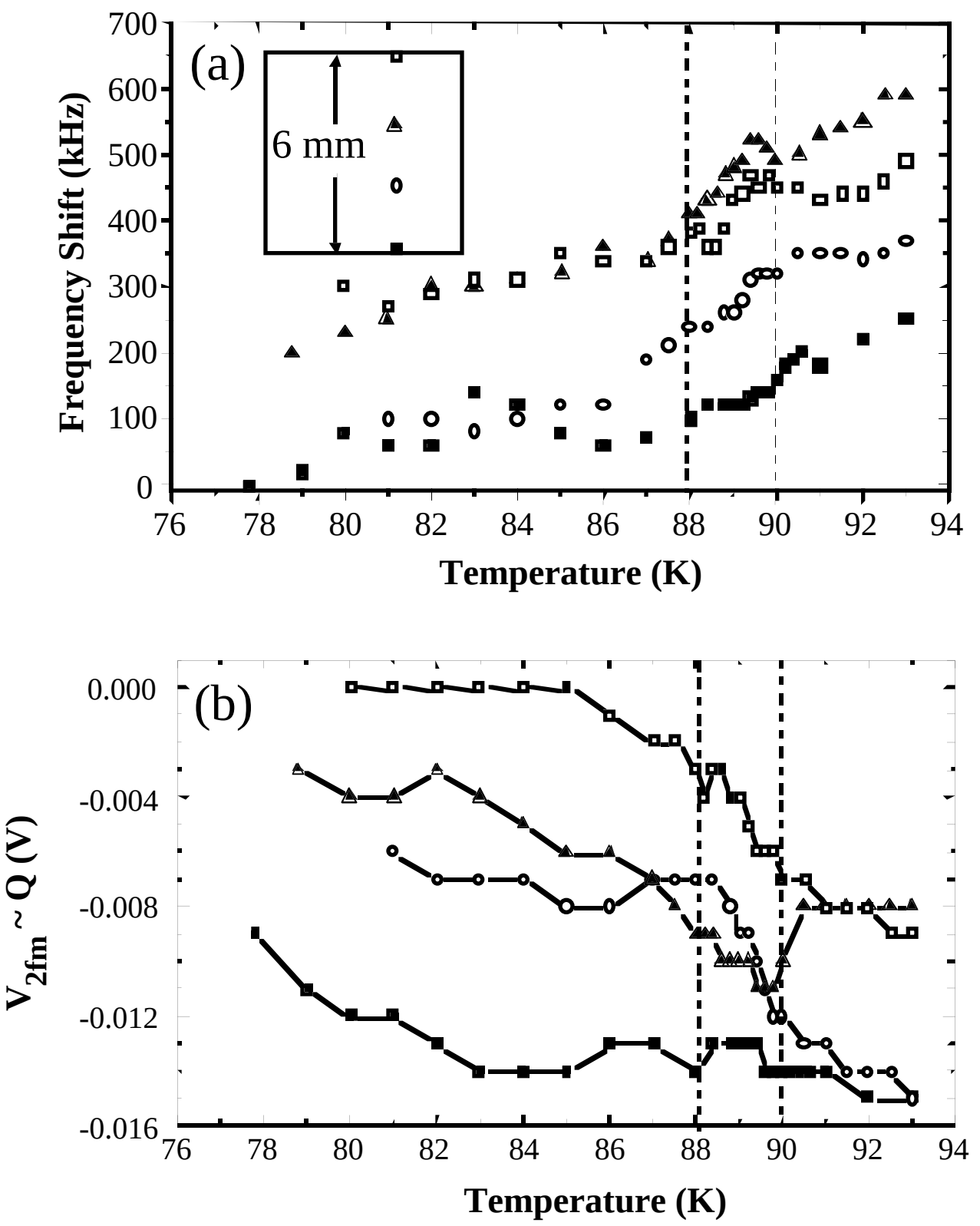

Fig. 8. Temperature dependence of $\Delta f$ and Q at several positions on a $YBa_2Cu_3O_{7-\delta}$ thin film. The curves have been displaced for the sake of clarity. The inset to the upper panel depicts the YBCO thin film indicating the positions at which the corresponding curves were measured.

We note that the curves for the frequency shift exhibit a similar overall slope. This is due to the drift of the source frequency with time. Care was taken to ramp the temperature in a similar fashion, resulting in the same slope for all curves. However, we note that the slope changes abruptly between about 88 K and 90 K for all curves, due to the rapidly changing surface impedance. Disregarding the overall drift, we would anticipate a constant $\Delta f$ below $T_C$, since the surface impedance drops below the sensitivity limit of the microscope. Above $T_C$, $\Delta f$ becomes constant as well due to the small temperature dependence of $Z_S$ in the normal state. Only within the transition region should a considerable change be observed.

A similar behavior can be seen for the quality factor Q, plotted in Fig. 8b. However, the changes are smaller and the scatter on the curves is somewhat larger. The superconducting transition of this film was checked independently by dc-resistivity measurements, yielding a value of 89.8 ± 0.2 K, in reasonable agreement with Fig. 8. In general, one expects the dc-transition to be sharper than a transition at microwave frequencies, consistent with the somewhat broader transition observed in Fig. 8.

Even though some work needs to be done to improve the signal-to-noise ratio, we can see that there are significant qualitative differences between the curves taken at different locations. For instance, the curves for both $\Delta f$ and Q at the position indicated by the open circle, start deviating at a temperature which is considerably higher than 88 K. This shows that, as expected, $T_C$ at this position is somewhat higher than at the edges of the film, an observation that would have been difficult to make with ordinary ways of measuring $T_C$.

## V. Microscope Limitations

### *A. Imaging Patterned Materials*

Although operating in the near-field limit offers the opportunity to improve the spatial resolution to dimensions as small as $\lambda/10^4$, one needs to be careful when measuring features that have dimensions comparable to the guided wavelength used. This is illustrated in Fig. 9, in which we show the results of imaging patterned Cu strips with varying dimensions on a printed circuit board. A photograph of the sample itself is shown in Fig. 9(c). The lines are of varying length and width, and were all scanned at a microscope frequency of 7.5 GHz.

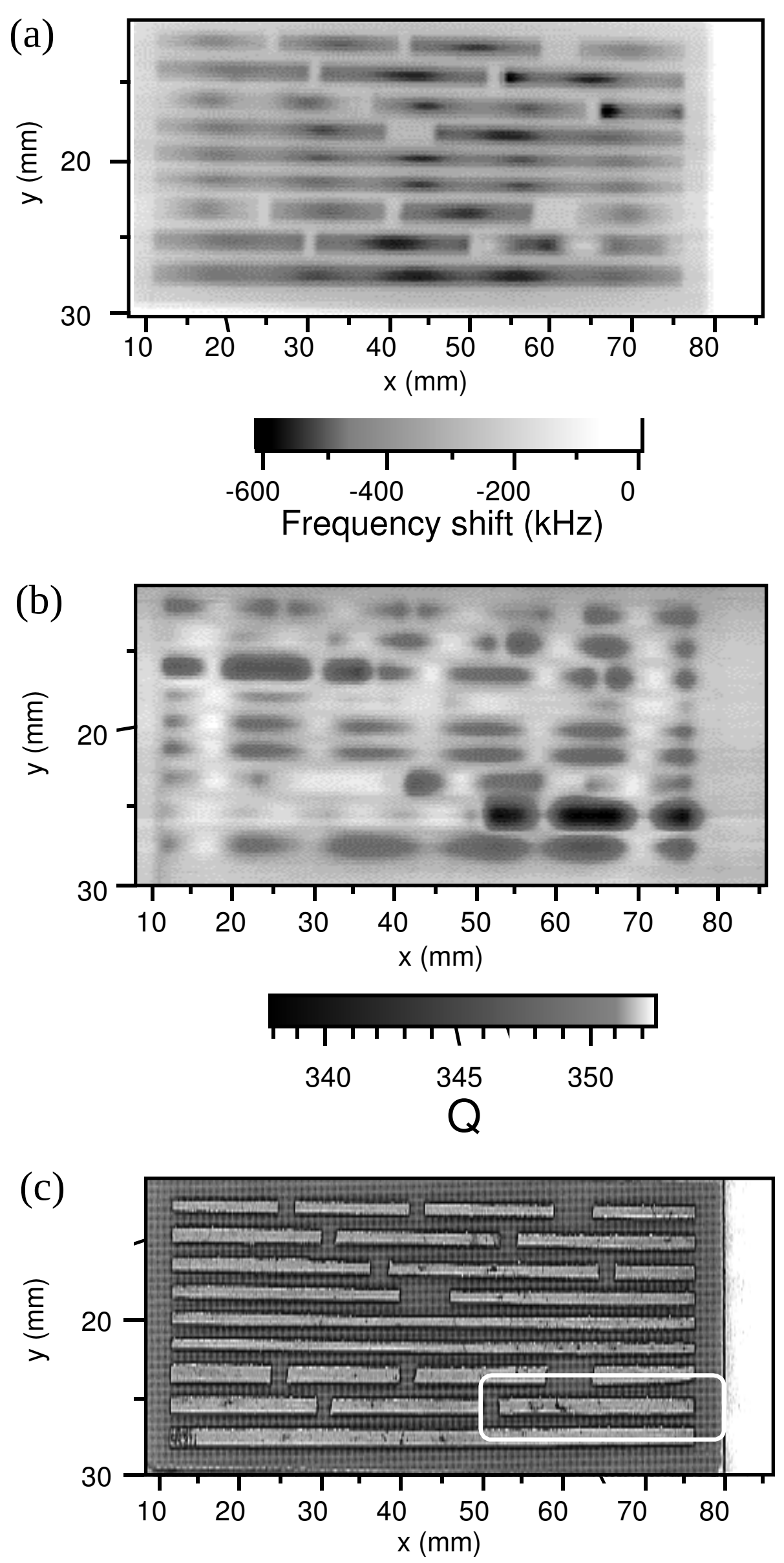

Fig. 9. a) Frequency shift image of a series of Cu-strips with different widths and lengths. (b): Corresponding Q image. (c): Photograph of the Cu-strips patterned on a printed circuit board. Note that the images have been elongated vertically for clarity.

Both the frequency shift image (Fig. 9(a)) and the Q image (Fig. 9(b)), are depicted. The probe size was 480 µm and the height of the probe above the sample was 200 µm. We see a rather distinct modulated pattern in both images, which is not merely an image of the metallic features of the sample seen in the photograph. The deviations from the expected behavior can be explained when we realize that most of the strips have a length which is similar to, or greater than, the guided wavelength of the microwaves used in imaging.

At the imaging frequency, the free-space wavelength is $\lambda_{air}$ = 4 cm, and the guided wavelength within the medium is about 2.5 cm [9]. As a result, some of the strips are excited into a resonant antenna-like mode, and display a standing wave pattern, instead of the expected uniform behavior. This is particularly clear for the strip highlighted in Fig. 9(c). This strip acts as a full-wave resonator, with minima in the Q at its two ends and in the middle. The guided wavelength which one deduces from this Q image is about 2.7 cm, very close to our estimate above.

We also note that the periodicity of the images in the inductive ($\Delta f$) and the absorptive (Q) part of the dielectric function are different. The Q shows a minimum at the ends of the strips, whereas the frequency shift peaks at the ends, thereby having one more maximum along the strip. In summary, this shows that the microwave microscope has difficulty in properly imaging patterned samples, or samples with edges, whenever resonant or antenna-like modes can be excited.

### *B. Other Limitations*

Another important issue is the drift of the microwave source center frequency with time, mentioned in Sec. IV.D. This has the effect of slowly changing the frequency shift signal during a scan, and can lead to an artificial "tilt" of the sample. It is necessary to use microwave sources with a high precision time base, to warm up the source for an extended period of time, and to limit the duration of the scan. However, we can also correct for the drift by regularly measuring the unperturbed microscope resonant frequency throughout the scan.

## VI. Conclusions

We have demonstrated a simple and effective microwave imaging system which allows us to image the response of a material to microwave electromagnetic fields. The technique has nearly continuous frequency coverage from 50 MHz to 50 GHz, has a spatial resolution which ranges from the micron to the millimeter level, and can examine samples up to 6" square at room temperature, and 2" square at cryogenic temperatures. Our cryogenic system can image objects at variable temperatures from 4.2 K up to room temperature, permitting the examination of superconducting microwave materials and devices in their natural operating state.

More generally, we have demonstrated quantitative measurements of material properties with these near field microscopes. In their present form, the microscopes have a sensitivity to sheet resistance of 0.6 Ω/sq. at 100 Ω/sq., and can measure reliably a minimum sheet resistance value of 0.4 Ω/sq. These numbers can be improved through the development of higher-Q coaxial resonators. The microscope is also sensitive to topography, having 55 nm height resolution at 30 µm probe/sample separation above a conducting sample. We have also found that the microscope is sensitive to spatially varying dielectric properties, and this will be the focus of future publications. The bandwidth, temperature range, spatial resolution capabilities, and quantitative imaging modes of this microscope make it very promising for superconducting microwave material characterization.

## Acknowledgments

We wish to thank Alberto Piqué for the YBCO wafer, Ken Stewart for the infrared polarizer grid, Rick Newrock for construction of the cryogenic microwave microscope probe, and D. Strachan for the YBCO film.